\documentclass[aps,pra,superscriptaddress,twocolumn,10pt]{revtex4-1}

\usepackage[utf8]{inputenc}
\usepackage[version=3]{mhchem}
\usepackage{amsmath}
\usepackage{amsfonts}
\usepackage{hyperref}
\usepackage{braket}
\usepackage{lipsum}
\usepackage{url}
\usepackage[draft]{graphicx}
\usepackage{lipsum}
\usepackage{color}
\usepackage{float}
\usepackage[export]{adjustbox}
\usepackage{wrapfig}
\usepackage{array,multirow}
\usepackage{tabularx}
\usepackage{bm}
\usepackage{comment}
\usepackage{relsize}

\makeatletter
\renewcommand\frontmatter@abstractwidth{\dimexpr\textwidth-1in\relax}
\makeatother
\usepackage[normalem]{ulem}
\usepackage{color}
\definecolor{my_red}{rgb}{.7,0,0}
\definecolor{my_green}{rgb}{0,.7,0}
\definecolor{my_blue}{rgb}{0,0,.7}
\definecolor{my_orange}{rgb}{1,.5,0}
\definecolor{my_purple}{rgb}{.5,0,.5}

\begin{abstract}

We compare the high-harmonic-generation (HHG) yield driven by a mid-infrared laser  in three  organic ring-shaped molecules, calculated using time-dependent density-functional theory (TDDFT). We average the yield over the relative orientation of the molecules and the linearly-polarized, 1825 nm driving laser pulse in order to compare to experimental spectra obtained by Alharbi {\it et al.}, Phys. Rev. A {\bf 92}, 041801 (2015). We find that the raw TDDFT-calculated HHG yield in cyclohexane (CHA) is strongly overestimated compared to those of benzene and cyclohexene, and that this can be attributed to unphysically large contributions from CHA orbitals lying well below the highest-occupied molecular orbital. We show that implementing a simple orbital-resolved scaling factor, which corrects the yield of the tunneling ionization contribution to the first step in the HHG process, leads to much better comparisons with experimental results. Our results are encouraging for the use of TDDFT in systematic computations of HHG in large molecules.  

\end{abstract}

\begin{document}

\author{Stephanie~N.~Armond}
\thanks{These two authors contributed equally to this work.}
\affiliation{Department of Physics and Astronomy, Louisiana State University, Baton Rouge, LA 70803, USA}

\author{Kyle~A.~Hamer}
\thanks{These two authors contributed equally to this work.}
\affiliation{Department of Physics and Astronomy, Louisiana State University, Baton Rouge, LA 70803, USA}

\author{Ravi~Bhardwaj}
\affiliation{Department of Physics, University of Ottawa, Ottawa, Ontario K1N 6N5, Canada}

\author{Francois Mauger}
\affiliation{Department of Physics and Astronomy, Louisiana State University, Baton Rouge, LA 70803, USA}

\author{Kenneth~Lopata}
\affiliation{Department of Chemistry, Louisiana State University, Baton Rouge, LA 70803, USA}
\affiliation{Center for Computation and Technology, Louisiana State University, Baton Rouge, LA 70803, USA}

\author{Kenneth~J.~Schafer}
\affiliation{Department of Physics and Astronomy, Louisiana State University, Baton Rouge, LA 70803, USA}

\author{Mette~B.~Gaarde}
\affiliation{Department of Physics and Astronomy, Louisiana State University, Baton Rouge, LA 70803, USA}

\title{Quantitative comparison of TDDFT-calculated HHG yields\\ in ring-shaped organic molecules}

\date{\today}
\maketitle


\section{Introduction}

High-harmonic spectroscopy (HHS) has emerged as a powerful tool for studying the structure and dynamics of molecules \cite{lein2003, marangos2008, smirnova2009, marangos2016}. HHS is implemented by characterizing the variation of the strong-field-driven high-harmonic generation (HHG) spectrum of a molecule  with the parameters of the laser field and the molecular configuration, and is typically combined with high-level theoretical calculations. The extraction of structural and/or dynamical information from HHS is possible because of the exquisite spatial and temporal resolution that is inherent in the  extremely nonlinear HHG process \cite{Nobel2023, schafer1993, corkum1993, Lewenstein1994}. Examples of HHS in molecules include studying vibrational dynamics \cite{lein2003}, characterizing structural features such as  two-center interference minima in a range of small molecules \cite{kanai2005, vozzi2005, hamer2021, tuthill2022}, and more recently imaging attosecond charge migration in organic molecules \cite{kraus2015, tuthill2020, he2022, hamer2022, hamer2023}. 

Accurate, ab-initio calculations of HHG spectra from molecules driven by mid-infrared (MIR) laser fields are very challenging, due to the large number of interacting electrons and the extreme nonlinearity of the HHG process. For example, since the HHG process involves both tunneling ionization and acceleration of the ionized electron far away from the molecular ion \cite{schafer1993, corkum1993}, it is necessary for a calculation to be accurate at both short  and long range. 
A number of calculations have been done with approximate methods based on the strong-field approximation, in which the HHG process is explicitly split up into its tunneling, acceleration, and recombination parts \cite{Zhou2005, Madsen2006, smirnova2009, lin2009}. For atoms and small molecules, explicitly correlated calculations based on multi-configuration methods have been successful in reproducing features in the HHG spectrum stemming from strong multi-electron correlation effects \cite{Pabst2013, Wahyutama2019, Bedurke2021, Ishikawa2024}. However, for larger molecules the only viable approach to systematic calculations of HHG spectra is time-dependent density-functional theory (TDDFT), in particular as implemented on a real-space grid that is large enough to contain the HHG process \cite{kohn1965, runge1984, Wardlow2016, hamer2021, coccia2022, Chu2024, Neufeld2024}. A challenge with TDDFT calculations is the choice of exchange-correlation (XC) potential in terms of both accuracy and computational cost. For example, it is well known that ground-state DFT calculations often underestimate orbital binding energies in molecules \cite{casida2000, lemierre2005, mack2013, Wardlow2016}, which for HHG calculations can lead to an artificial increase of the influence of lower-lying orbitals compared to that of the highest-occupied molecular orbital (HOMO) \cite{hamer2021}. There are a number of sophisticated hybrid-type XC potentials available that are known to yield excellent description of molecular structure and linear dynamics \cite{burke2012perspective, bursch2022best, laurent2013td, dreuw2003long, baer2010tuned}, as well as strong-field ionization \cite{sissay2016}. However, several studies have shown that these potentials can also introduce unphysical features into the HHG spectrum, and that a safer choice might in fact be using the local-density-approximation (LDA) potential \cite{mack2013, Neufeld2024}, in combination with a self-interaction-correction. Ultimately, progress in this area will require careful comparisons between different types of calculations, and between calculated and experimental results.

In this paper, we use grid-based TDDFT to compare the orientation-averaged HHG yield between three six-membered ring molecules: benzene (BNZ, C$_{6}$H$_{6}$), cyclohexene (CHE, C$_{6}$H$_{10}$), and cyclohexane (CHA, C$_{6}$H$_{12}$), driven by an MIR laser field. We then compare these results with experimental yields \cite{alharbi2015}, which showed that the HHG yields are positively correlated with the aromaticity of each molecule: BNZ $>$ CHE $>$ CHA. We find that the raw TDDFT output, computed using an LDA XC potential, including an average-density self-interaction correction (ADSIC), overestimates the relative yield of CHE and especially CHA, such that the ordering of the calculated HHG yields differs from that of the experimental yields. We show that the increased HHG yield in CHE and CHA is due to unphysically-large contributions from lower-lying orbitals (especially the HOMO-2 in CHA) whose  ionization potentials are underestimated by the TDDFT calculations. Finally, we show that a simple correction of the HHG yields, by a Kohn-Sham-orbital-resolved scaling factor that adjusts the yield of the ionization step in the HHG process, leads to substantially better comparison with experimental results, and in particular to a strong depression of the CHA yield compared to the other two molecules. The correction factor rescales the tunnel ionization rates in our calculations using experimentally measured ionization potentials in place of the DFT-predicted ionization potentials. This is thus in the spirit of the three-step model in which the HHG process is explicitly split up into its tunneling, acceleration, and recombination parts \cite{smirnova2009, lin2009}. 

The manuscript is arranged as follows: Section II describes our theoretical approach, detailing the process of TDDFT-computations of orientation-averaged HHG spectra using the Lebedev quadrature \cite{lebedev1975}. Section III presents the computed orientation-averaged HHG yields, and describes the process of rescaling the orbital-resolved dipole signals in order to quantitatively compare HHG yields between different molecules. Lastly, in Section IV, we provide a summary and discuss the prospects of our results. Unless otherwise stated, atomic units (a.u.) are used throughout the paper.


\section{Methodology}


\subsection{TDDFT Simulations} \label{sec:Method:TDDFT_simulations}

For each molecule, we start by calculating the ground state using density-functional theory (DFT). We use a LDA-with-ADSIC XC potential~\cite{bloch1929, dirac1930, perdew1981, marques2012, legrand2002}, and we employ pseudopotentials to describe the screening of the innermost electrons \cite{trouiller1991, froyen1982, soler2002}. All simulations are done using the real-space, grid-based software package Octopus \cite{andrade2012, andrade2015}. The molecule is centered inside a rectangular box that is 50 $\times$ 50 a.u. wide, and 90 or 110 a.u. long, depending on the laser intensity. 

We then perform a time-dependent calculation of the laser-driven dynamics, solving the Kohn-Sham equations to calculate the time-dependent electron density \cite{kohn1965, runge1984}. We keep the nuclear geometry frozen in all calculations \footnote{In \cite{Wardlow2016}, nuclear motion  was found to influence the details of the HHG spectrum, but not the overall yield}. The 1825~nm laser field is ramped up linearly over two optical cycles to a peak intensity of 28 or 45 TW/cm$^2$, and then has a constant amplitude for five cycles. The laser field is linearly polarized along the $x$-direction, and the molecule is rotated in the box to simulate different relative orientations.  The longest dimension of the box is along the laser polarization and is scaled to the semi-classical electron quiver radius \cite{schafer1993, corkum1993}. This allows us to effectively filter away the long-trajectory contribution to the HHG spectrum using a complex absorbing potential (CAP), as was done in \cite{hamer2021}. We use a $\sin^{2}$ CAP that extends 10 a.u.\ from each edge of the box, and the spatial grid has a 0.3 a.u.\ spacing in all directions. 

\begin{figure}[tb]
    \centering
    \includegraphics[width=0.95\columnwidth]{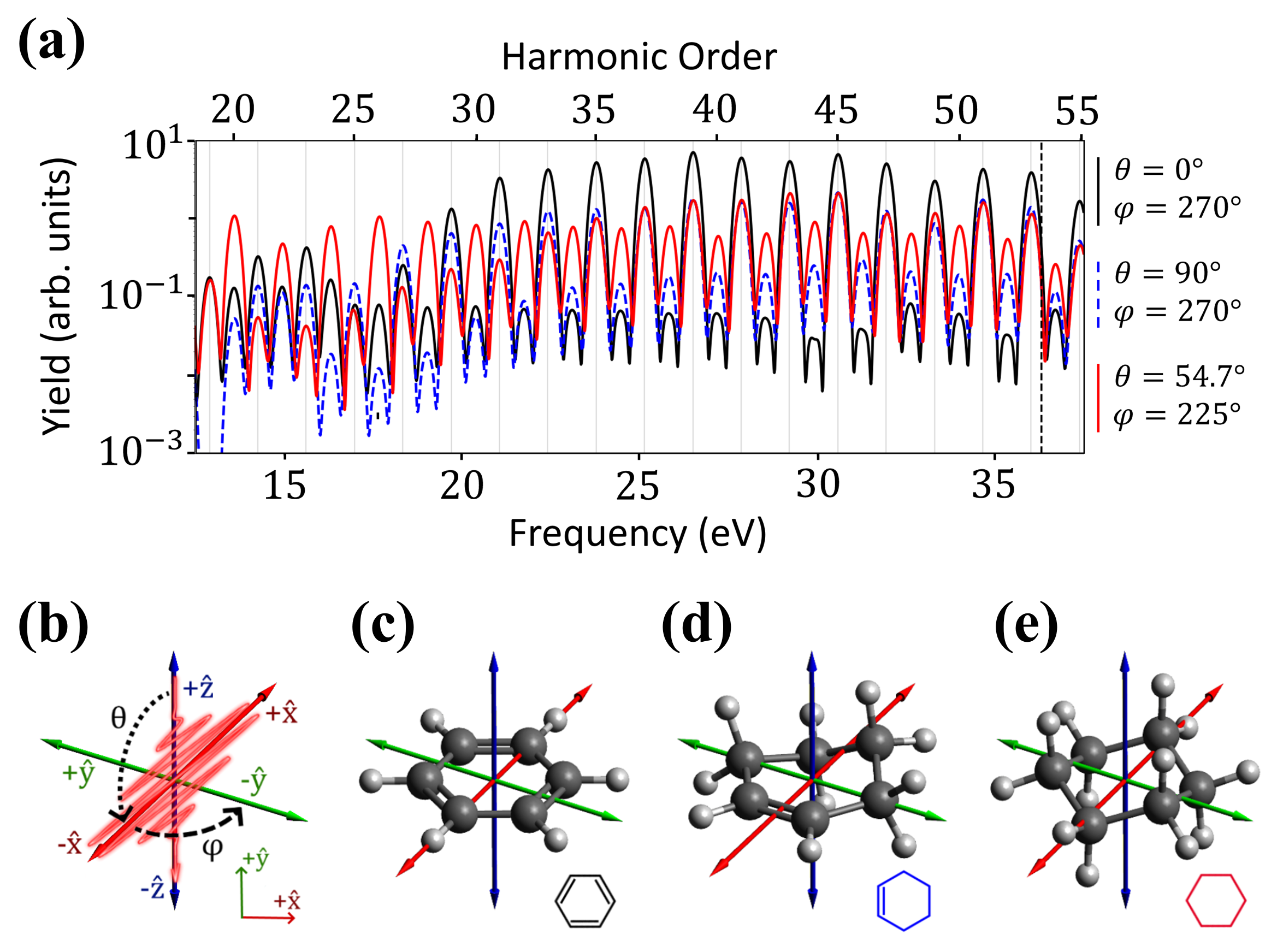}
    \caption{(a) HHG yield in CHE for three different orientations of the molecule relative to the laser polarization, driven by a 1825~nm, 28 TW/cm$^2$ laser pulse. The dotted black vertical line marks the semi-classical prediction for the cutoff energy. 
    (b) Definition of the rotation angles $\theta$ and $\phi$ for the molecule relative to the fixed laser polarization along the $x$-axis, with the $\theta=0, \phi=0$ orientation illustrated in the molecular diagrams in (c) BNZ, (d) CHE, and (e) CHA.
    }
    \label{fig:che_angles}
\end{figure}

For each orientation of the molecule, characterized by angles $\theta$ and $\phi$, the harmonic spectrum is calculated from the windowed Fourier transform of the time-dependent dipole acceleration: 
\begin{equation}\label{eq:spectrum}
    S[\theta, \phi](\omega)=\sum_{j=x,y} {\left|\mathcal{F} \left(W(t) \cdot a_{j}[\theta, \phi](t)\right)  \right|}^2
\end{equation}
where the window $W(t)$ is a $\cos^{2}$ function that spans the final five cycles of the laser field. We have chosen the laser field to propagate along the $z$-direction and therefore exclude the $a_z(t)$ contribution to the harmonic yield since it would not lead to light propagating in the forward direction. 

Fig.~\ref{fig:che_angles}(a) illustrates the HHG spectrum in CHE for three different molecular orientations, for an intensity of 28 TW/cm$^2$, with the angles $\theta$ and $\phi$ illustrated in panel (b). The yield and the shape of the spectrum clearly varies strongly as a function of orientation. This has been observed in experiments and calculations \cite{lein2003, yun2017, tuthill2022} for a number of molecules, and can often be understood in terms of the shape of the HOMO of the molecule. For example, the harmonic yield will be strongly suppressed if the laser is polarized along a nodal plane of the HOMO \cite{madsen2007}. The prominent even harmonics in some of the spectra arise due to the lack of inversion symmetry for some CHE orientations. In a gas of molecules with random orientations, inversion symmetry is present at the macroscopic level and the even harmonics will cancel out. We will discuss how to average over the molecular orientation below. 

\subsection{Angle Averaging using the Lebedev Quadrature} \label{sec:Method:lebedev}

\begin{figure*}[ht]
    \centering
    \includegraphics[width=\linewidth]{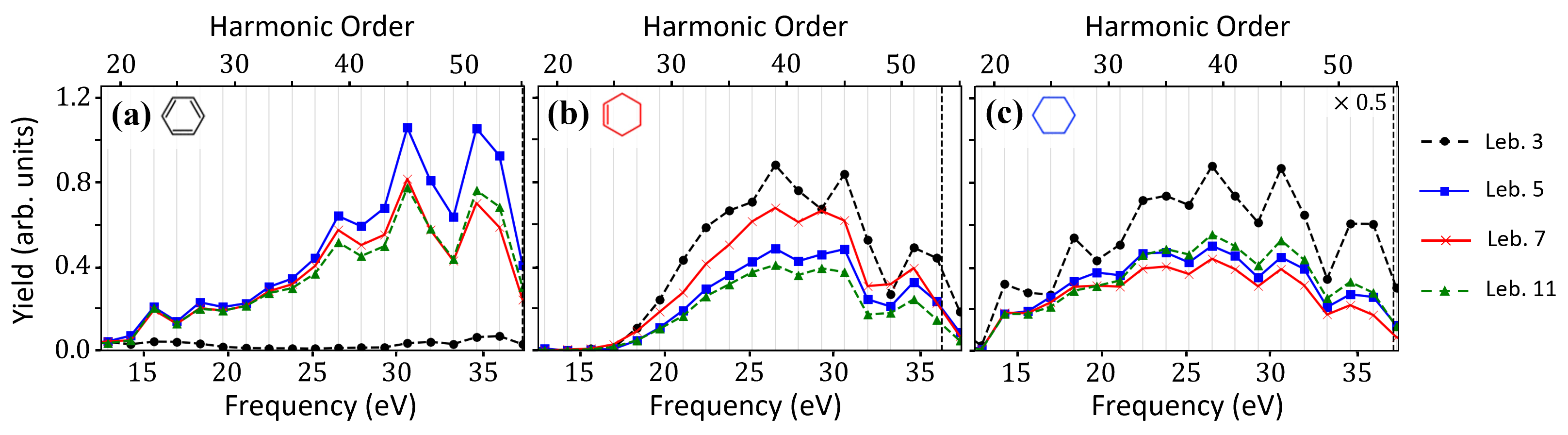}
    \caption{Convergence of orientation-averaged HHG yields, using Lebedev 3, 5, 7, and 11 grids, for (a) BNZ, (b) CHE, and (c) CHA. The peak intensity of the 1825~nm laser field is 28~TW/$\text{cm}^{2}$, and the semi-classical cutoff energy is marked with a dashed, vertical line.} 
    \label{fig:lebedev_convergence}
\end{figure*}

In order to compare our TDDFT-calculated HHG yields to experimental yields measured in unaligned molecular gases, we have to coherently average the harmonic yield over the angular orientation on the unit sphere $\mathbb{S}^{2}$:
\begin{equation}\label{eq:iso_spec_integral}
    S_{tot}(\omega) = \sum_{j=x,y} {\left| \int_{\mathbb{S}^{2}} \tilde{a}_{j}[\theta, \phi](\omega) \sin{\theta}\, \mathrm{d}\theta\, \mathrm{d}\phi \right|}^2
\end{equation}
where $\tilde{a}_{j}[\theta, \phi](\omega)$ is the windowed Fourier transform of the dipole acceleration at a given molecular orientation. The rotation angles $\theta$ and $\phi$ are defined in Fig.~{\ref{fig:che_angles}}(b). Given that each HHG calculation is computationally expensive ($\approx$ 2,500 CPU hours on a state-of-the-art supercomputer system), we need to make the angular averaging as efficient as possible. We do this by employing the Lebedev quadrature sampling of $\mathbb{S}^{2}$ \cite{lebedev1975}, such that Eq.~{\ref{eq:iso_spec_integral}} above becomes a sum:
\begin{equation}\label{eq:iso_spec_sum}
    S_{tot}(\omega) \equiv \sum_{j=x,y} {\left| \sum_{i=1}^{N} w_{i}\, \tilde{a}_{j}[\theta_{i}, \phi_{i}](\omega) \right|}^2
\end{equation}
where the number of points $N$ on $\mathbb{S}^{2}$ is determined by the order $k$ of the Lebedev grid, and the weights $w_{i}$ and grid points $(\theta_{i}, \phi_{i})$ are chosen to have both octahedral rotation and inversion symmetry. 

To further reduce the number of simulations that we have to perform, we take advantage of the so-called dynamical symmetries \cite{ceccherini2001, bunker2006} of the molecules in their interaction with the periodic laser field. As an example, rotating a BNZ molecule by $\pi$ radians in the plane of the molecule results in the exact same lab-frame ground-state electron density. This symmetry is then preserved when the field is turned on, leading to identical harmonic signals in all spatial directions. If the densities of two orientations are identical after one is rotated by $\pi$ radians around the $x$-axis (the laser polarization axis), the harmonic signal along the $x$-axis will be identical. However, when adding up the contributions from these two orientations, we have to take into account that both perpendicular components have opposite signs because of the sign change of the electric field, $E(t+T/2)=-E(t)$. Similarly, if the densities are identical under a $\pi$ rotation around the $y$- ($z$-) axis, then the contribution from the rotated orientation is added with a sign change on the $y$- ($z$-) component of the acceleration, when taking into account the periodicity of the electric field. In this way, we reduce the number of simulations required by a factor of three to five, depending on the molecular symmetry. More details about the number of calculations required for a particular Lebedev order are provided in Appendix A. 

The orientations of the three molecules for $\theta=0, \phi=0$ are shown in Fig.~\ref{fig:che_angles}(c-e). Finally, we note that we have used the ``chair'' conformer for CHA in all simulations, given the abundance ($>$99\% \cite{ramsay1947}) of this conformer at room temperature. 

\section{Results}

We begin by showing the convergence of the total harmonic spectrum with respect to the angular averaging, in Figure{~\ref{fig:lebedev_convergence}}. For clarity, we plot the yield for each odd harmonic integrated over $0.5\omega_{L}$ around each peak, where $\omega_{L} = 0.68\ \text{eV}$ is the laser frequency. We show spectra integrated in Lebedev order 3, 5, 7, and 11 for all three molecules, with the Lebedev 11 grid consisting of 50 points on $\mathbb{S}^{2}$ in total (without accounting for the dynamical symmetries). The angles included in these particular orders build on each other, such that all order-3 angles are contained in the order-5 grid, and so on. Comparing these results with Lebedev 9 or 13 would require a new set of calculations as some of the angles sampled in these orders are different from the ones used here. 

\begin{figure}[tb]
    \centering
    \includegraphics[width=\linewidth]{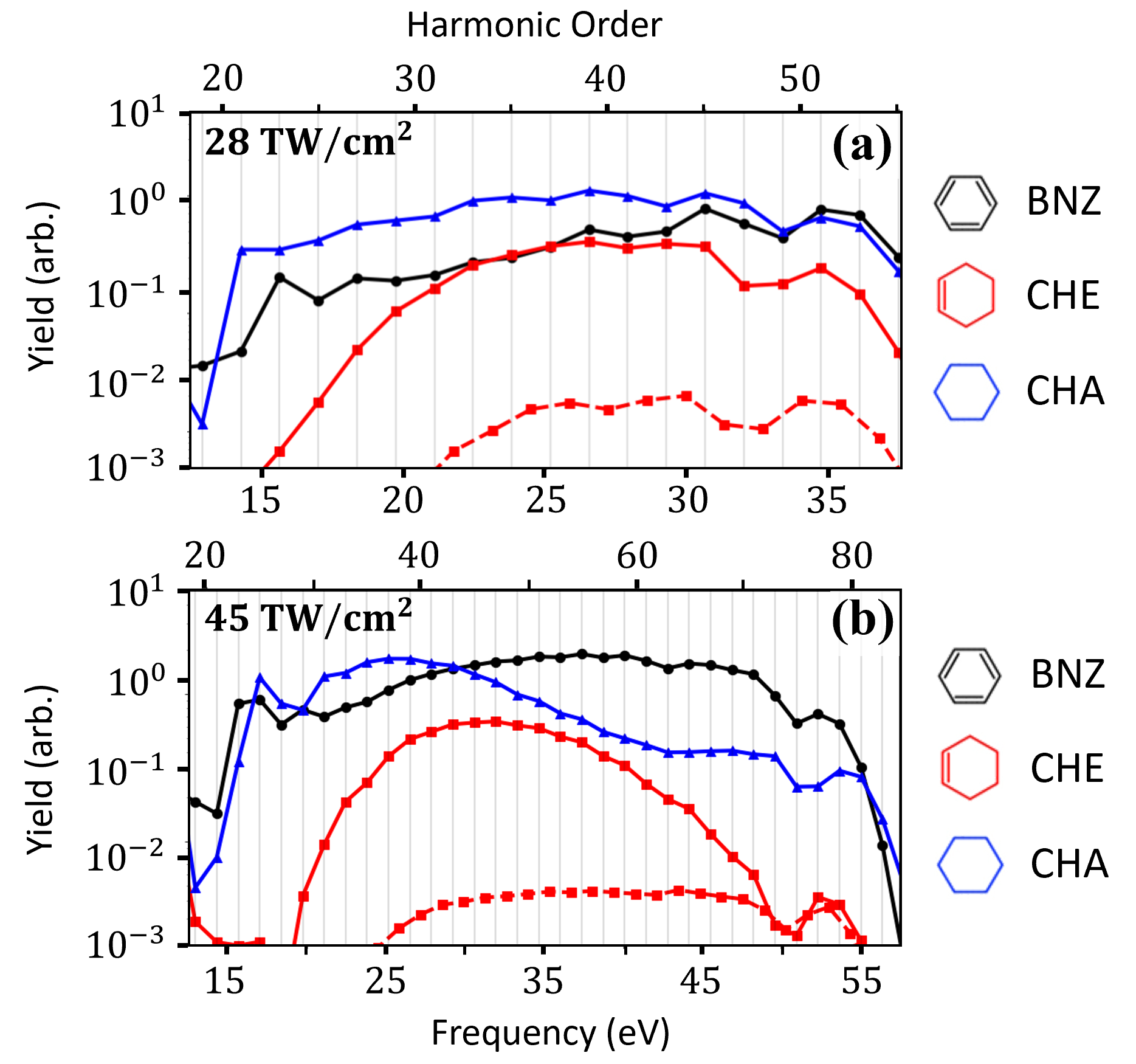}
    \caption{Comparison of HHG yields in BNZ, CHE, and CHA for laser intensities of (a) 28~TW/$\text{cm}^{2}$ and (b) 45~TW/$\text{cm}^{2}$ at ${\lambda} = 1825$~$\text{nm}$. Note that the frequency range is different in panels (a) and (b). The red dashed line shows the even-harmonic yield in CHE.}
    \label{fig:unweighted_spectra}
\end{figure}

For all three molecules, Fig.~\ref{fig:lebedev_convergence} shows that already the Lebedev 5 results are remarkably well converged, especially with respect to the shape of the spectra. The yield in BNZ is converged for Lebedev 7, whereas for CHE and CHA the yields change by less than a factor of two between Lebedev 7 and 11 (which corresponds to doubling the number of angular grid points). This bodes well for calculations in which one is interested in the overall shape and yield of molecular harmonics: for most molecules Lebedev 5 or 7 averaging is likely sufficient. In the remainder of this manuscript, we show results averaged using Lebedev 11. A measure of the convergence of the Lebedev 11 calculations can be seen in the yield of the spectrally resolved even harmonics in the angularly averaged CHE spectra shown in Fig.~\ref{fig:unweighted_spectra}. In a perfectly averaged spectrum, the even harmonic yield should be zero -- the fact that it is around 5\% of the odd harmonic yield suggests that the overall angular convergence is within 50\%. 

\begin{figure*}[!t]
    \centering
    \includegraphics[width=\linewidth]{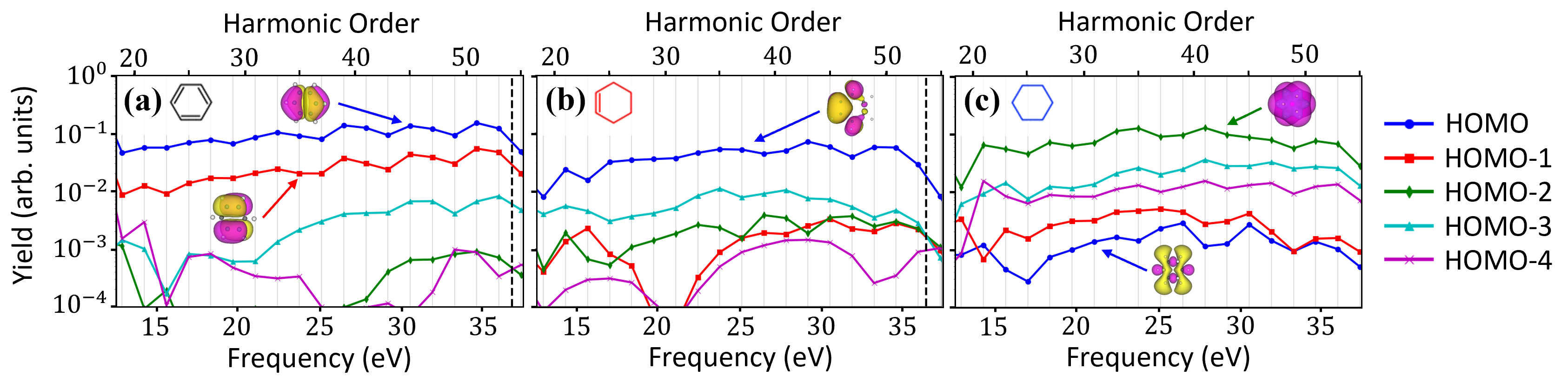}
    \caption{Orientation-averaged HHG spectra resolved into individual orbital contributions for BNZ (a), CHE (b), and CHA (c), for an intensity of $28$~TW/$\text{cm}^{2}$. The insets depict isosurfaces for the orbitals with the dominant contributions for each molecule. Isosurfaces are taken at each molecule's ${\theta} = 0^{\circ}$, ${\phi} = 0^{\circ}$ orientation.}
    \label{fig:orbital_resolution}
\end{figure*}

We are now ready to directly compare the HHG yields in the three molecules, as was done in the experiments by Alharbi and collaborators \cite{alharbi2015}. This is shown in Fig.~\ref{fig:unweighted_spectra} for two different laser intensities: 28~TW/$\text{cm}^{2}$ and 45~TW/$\text{cm}^{2}$. In \cite{alharbi2015},  the ratio of the BNZ:CHE yields was found to vary between 5 and 15, for different intensities ranging between 45 and 86 TW/cm$^2$. Our BNZ:CHE ratios are in reasonable agreement with this, although substantially smaller for the lower intensity. In contrast, the calculated HHG yields in CHA are much bigger than those of \cite{alharbi2015}: for the lower intensity, the CHA yields are higher than the BNZ yields except in the cutoff region, and at the higher intensity the CHA yields are in between the BNZ and CHE yields across most of the plateau region. In contrast, the experimental CHA yields were much lower than both the CHE and the BNZ yields, with a BNZ:CHA ratio of approximately 50-100. We note here that, similar to the conclusion in \cite{alharbi2015}, we find that the differences in yield between the molecules cannot be accounted for with the difference in the total ionization rates. For example, by monitoring the time-dependent electron density absorbed at the edges of the box, we estimate that the orientation-averaged ionization rate is greater in CHE than in BNZ, opposite from the trend seen in Fig.~{\ref{fig:unweighted_spectra}}. 

We can investigate the origin of the disagreement between the calculated and measured yield ratios by looking at how the yield for the three molecules is composed of contributions from different Kohn-Sham orbitals \cite{hamer2021}. This is shown in Fig.~\ref{fig:orbital_resolution}, which contains the orbital-resolved orientation-averaged yields for BNZ, CHE, and CHA in panels (a-c). For each molecule, we also show the orbital shapes of the dominant contributions as insets. Although the Kohn-Sham orbitals evolve in time when the laser field is on (somewhat analogously to the field-dressed orbitals in the real system), they retain their initial symmetries and can therefore provide a reasonable measure of the ``true'' orbital-resolved yield. The orbital-resolved contributions are calculated from the Fourier transform of the time-dependent dipole moment (as opposed to the acceleration shown above), as the orbital-resolved dipole moments are what we have easy access to from Octopus output files. All dipole-based spectra have been multiplied by a factor of $\omega^4$ for direct comparison with the acceleration-based spectra. 
The agreement between the acceleration and the dipole yield is good, as shown in more detail in  Appendix B, with the dipole yield being slightly less smooth when averaging over angle. 

\setlength{\tabcolsep}{4pt}
\renewcommand{\arraystretch}{1.25}
\newcolumntype{?}[1]{!{\vrule width #1}}
\newcolumntype{C}{wc{8mm}}
\begin{table*}[htb]
    \centering
    \begin{tabular}{?{0.5mm}c?{0.5mm}C|C|C?{0.5mm}C|C|C?{0.5mm}C|C|C?{0.5mm}}
        \cline{2-10}
        \multicolumn{1}{c?{0.5mm}}{} & \multicolumn{3}{c?{0.5mm}}{BNZ ($\text{C}_{6}\text{H}_{6}$)} & \multicolumn{3}{c?{0.5mm}}{CHE ($\text{C}_{6}\text{H}_{10}$)} & \multicolumn{3}{c?{0.5mm}}{CHA ($\text{C}_{6}\text{H}_{12}$)}\\
        \hline
        Orbital & $I_{p}^{\text{exp}}$ & $I_{p}^{\text{thy}}$ & $\eta_{ADK}$ & $I_{p}^{\text{exp}}$ & $I_{p}^{\text{thy}}$ & $\eta_{ADK}$ & $I_{p}^{\text{exp}}$ & $I_{p}^{\text{thy}}$ & $\eta_{ADK}$\\
        \hline \hline
        HOMO & \multirow{2}*{9.25} & \multirow{2}*{9.78} & \multirow{2}*{3.57} & 8.94 & 8.74 & 0.62 & \multirow{2}*{10.32} & 9.63 & 0.19\\
        \cline{1-1} \cline{5-7} \cline{9-10}
        HOMO-1 & & & & 10.7 & 9.88 & 0.13 & & 9.65 & 0.19\\
        \hline
        HOMO-2 & \multirow{2}*{12.71} & \multirow{2}*{10.97} & \multirow{2}*{0.01} & 11.3 & 10.15 & 0.05 & 11.28 & 9.94 & 0.04\\
        \cline{1-1} \cline{5-10}
        HOMO-3 & & & & 11.7 & 10.42 & 0.04 & \multirow{2}*{12.00} & \multirow{2}*{10.64} & \multirow{2}*{0.03}\\
        \cline{1-7}
        HOMO-4 & 12.08 & 12.61 & 4.16 & 12.8 & 11.84 & 0.08 & & & \\
        \hline
    \end{tabular}
    \caption{Experimental and calculated ionization potentials ($I_{p}^{\text{exp}}$ and $I_{p}^{\text{thy}}$, respectively) and resulting ADK ratios (see text) for the five highest-occupied Kohn-Sham orbitals in BNZ, CHE, and CHA. $I_{p}^{\text{thy}}$ is calculated as the orbital binding energy in the DFT ground-state calculation. The ADK ratios are computed for an intensity of $28$~TW/$\text{cm}^{2}$.}
    \label{tab:ips}
\end{table*}

Fig.~\ref{fig:orbital_resolution}(a) shows that in BNZ, the degenerate HOMO and HOMO-1 completely dominate the yield. This is not surprising given that the binding energy of HOMO-2 is several eV larger than that of HOMO/HOMO-1, as will be discussed in more detail below. Note that the yields of the HOMO and the HOMO-1 are different; since BNZ has a six-fold symmetry in the plane of the molecule, the HOMO-1 is \emph{not} simply a $\pi/2$ rotation of the HOMO. The CHE yield is also dominated by the HOMO contribution, with some contributions from lower-lying orbitals. However, the CHA yield is dominated by the HOMO-2 (and lower) contribution(s), with the HOMO and HOMO-1 yields being much weaker. This is surprising given that the experimentally measured gap between the ionization potential ($I_p$) of the HOMO/HOMO-1 and the HOMO-2 in CHA is about 1 eV \cite{kimura1981}. It also indicates a potential cause of the too-high yield of the CHA HHG: when we look at our calculated ionization potentials $I_{p}^{\text{thy}}$ (taken to be equal to the orbital binding energy, which functions as the effective energy barrier for tunneling out of each orbital according to Koopmans' theorem) in Table~{\ref{tab:ips}}, we find that our $I_{p}^{\text{thy}}$ of the HOMO/HOMO-1 is 9.6 eV and that of HOMO-2 is 9.9 eV, compared to $I_{p}^{\text{exp}}$ of 10.3 eV and 11.3 eV in \cite{kimura1981}. Given the exponential sensitivity of the tunnel ionization rate to the ionization potential, this 1.3 eV difference between $I_{p}^{\text{thy}}$ and $I_{p}^{\text{exp}}$ is  important: it means that 
our DFT and TDDFT calculations underestimate the ionization potentials, and thus overestimate the contribution to the yield from these orbitals. The experimental and calculated values for $I_p$ for a number of orbitals in all three molecules is shown in Table \ref{tab:ips}.

Table \ref{tab:ips} also shows that while $I_{p}^{\text{thy}}$ of the HOMO in CHE is close to $I_{p}^{\text{exp}}$, DFT overestimates the $I_{p}^{\text{thy}}$ in BNZ and underestimates the $I_{p}^{\text{thy}}$'s of all the orbitals in CHA, in particular the HOMO-2 as discussed above. It is well known that DFT commonly underestimates orbital binding energies \cite{casida2000, lemierre2005, mack2013}, but given how sensitive the HHG yield is to the initial tunnel-ionization step of the process, this will skew the HHG yields in one direction or the other. In particular, the much-too-low $I_{p}^{\text{thy}}$ of the HHG-dominant HOMO-2 orbital in CHA is most likely the cause of the too-high HHG yield of CHA relative to the other molecules in our calculations as compared to the experiment \cite{alharbi2015}.

We next show that scaling the orbital-resolved yields by a simple tunnel-ionization-related correction factor will bring the total HHG yields into better agreement with the experimental results. According to the quantitative rescattering (QRS) formulation of the semi-classical three-step model \cite{lin2009, lin2010, mauger2016}, the HHG spectrum can be written as a product of the ionization rate, a generic function describing the propagation of the electron wave packet in the continuum, and the energy-dependent recombination dipole matrix element. In this framework, our rescaling corrects the ionization step. Specifically, we divide out the ADK tunnel ionization rate related to DFT-calculated $I_{p}^{\text{thy}}$, and multiply by that of the experimental $I_{p}^{\text{exp}}$:
\begin{equation}\label{eq:adk_ratio}
    \eta_{ADK} = \frac{\gamma_{\text{exp}}}{\gamma_{\text{thy}}}.
\end{equation}
We use the static-field ADK tunnel ionization rate $\gamma$ given by
\begin{widetext}
    \begin{equation}\label{eq:adk_rate}
        \gamma = (3.37 \times 10^{-3})\, n^{-9/2} \cdot \text{exp}\left[-\frac{2}{3 F\, n^{3}}\right] \cdot \left(\frac{10.87}{F\, n^{4}}\right)^{2n - 1.5}\ , \quad \text{with } \begin{cases} n = (I_{p} / 13.6)^{-1/2}\\ F = (5.35 \times 10^{-9}) \sqrt{I_{\circ}} \end{cases}
    \end{equation}
\end{widetext}
where the ionization potential $I_{p}$ is in eV, the laser intensity $I_{\circ}$ is in W/cm$^{2}$, and $\gamma$ is in units of $\text{s}^{-1}$ \cite{ammosov1986}. The resulting correction factors of Eq.~{\ref{eq:adk_ratio}} for each orbital are also shown in Table~\ref{tab:ips}. The reweighted HHG spectra are obtained by multiplying the orbital-resolved spectra of Fig.~{\ref{fig:orbital_resolution}} by the orbital-dependent ADK ratios in Table~\ref{tab:ips}. 

\begin{figure}[!t]
    \centering
    \includegraphics[width=\linewidth]{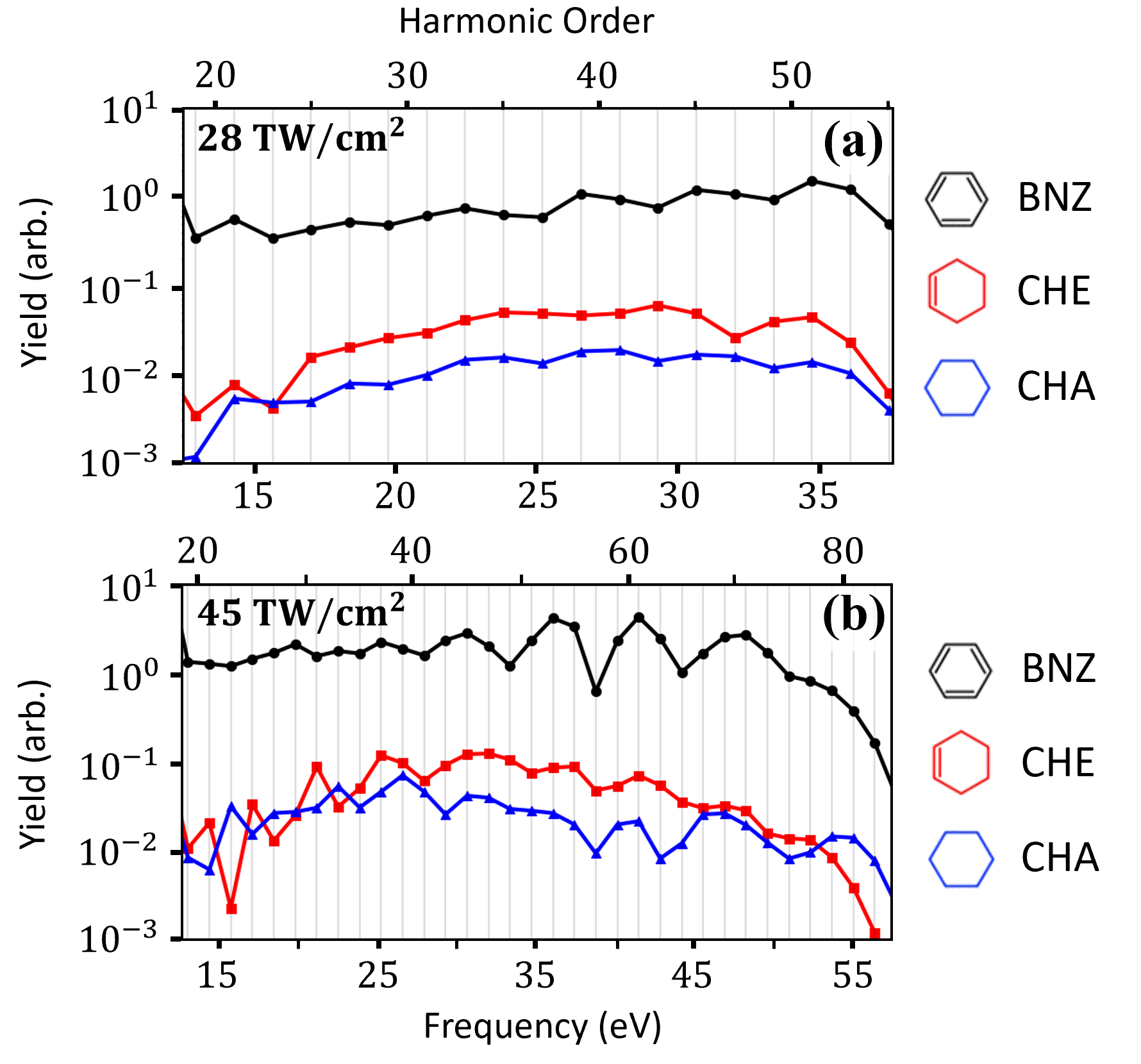}
    \caption{(a,b) Orientation-averaged HHG yields with the contributions from the highest-lying orbitals weighted by the ADK ratios in Table~{\ref{tab:ips}}, for two different intensities. (c) ADK-corrected HHG yield ratios between BNZ and either CHE or CHA, for both intensities.}
    \label{fig:corrected_spectra}
\end{figure}

We note here that more sophisticated approaches to calculating tunnel ionization rates, such as molecular ADK \cite{tong2002} and weak-field asymptotic theory \cite{tolstikhin2014, wahyutama2022}, all include a basic ADK-like exponential factor, along with factors that depend on properties of the electronic wave function. Our correction thus implicitly assumes the TDDFT-calculated orbital captures the correct spatial and temporal behavior of the true time-dependent wave function, even if the orbital binding energy is somewhat shifted. 
This is consistent with the general purpose of the Kohn-Sham orbitals \cite{baerends2009, baerends2013}. 
Furthermore, a number of studies have achieved good comparison between angle-dependent strong-field ionization TDDFT calculations and experimental results \cite{sissay2016, sandor2018}, as well as between few-femtosecond electron dynamics calculated with TDDFT and more explicitly correlated methods \cite{bruner2017}.

The orbital-scaled total spectra are shown in Fig.~\ref{fig:corrected_spectra} for the two different intensities. These results are more consistent with experiment, in that for both intensities the ordering of the yields (BNZ $>$ CHE $>$ CHA) within the harmonic plateau is consistent with Ref.~{\cite{alharbi2015}}, and the BNZ:CHE ratio is greater than the CHE:CHA ratio. Note that the oscillation in the 45 TW/cm$^{2}$ BNZ spectrum is due to interference between dipole signals from different orientations, but this oscillation does not affect the overall HHG yield -- see Appendix B for further details.

\begin{figure}[!t]
    \centering
    \includegraphics[width=\linewidth]{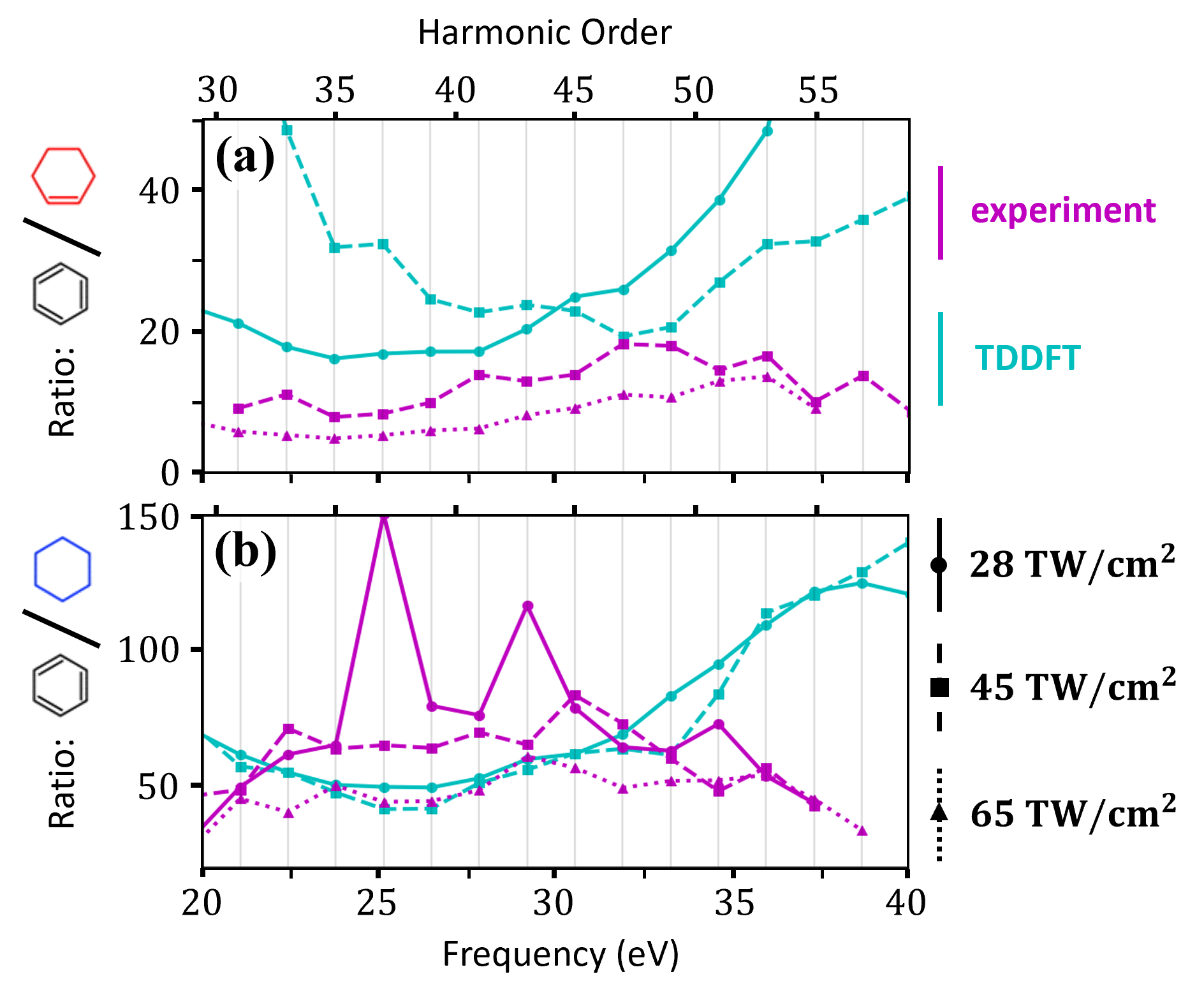}
    \caption{Comparison between experimental results (purple) \cite{alharbi2015} and our results using TDDFT (cyan), for the ratios between (a) BNZ and CHE, and (b) BNZ and CHA, for various intensities.}
    \label{fig:ratios}
\end{figure}

The ratios of the BNZ:CHE and the BNZ:CHA yields are shown in Figure~{\ref{fig:ratios}}(a) and (b), respectively. The purple curves are from experiment, some of which were published in \cite{alharbi2015}, while the cyan curves are taken from our TDDFT simulations. The different linestyles represent different laser intensities, as indicated by the legend to the right of panel (b). To better see the trend in the ratios, we smooth out the oscillations in the respective yields by showing a running average over five consecutive harmonic orders in all theory curves. The correction factor has only changed the BNZ:CHE ratio moderately, and it now varies between 10 and 50 for much of the plateau and then diverges in the cutoff region. In contrast, the BNZ:CHA ratio has changed drastically and now varies between approximately 50 and 150, similar to the experimental results. We note that the correction of the tunneling ionization contribution to the HHG yield does not alter the cutoff energies, which will be determined by the uncorrected $I_p$ values. This means that the corrected ratios are less meaningful in the cutoff region. 

The findings shown in Figs.~{\ref{fig:corrected_spectra}} and \ref{fig:ratios} are in much better agreement with experimental results than the original results of Fig.~{\ref{fig:unweighted_spectra}}. In particular the CHA yield relative to the other two molecules is much better, which is no surprise given that its lower-lying orbital $I_p$'s were the most offset from the experimental values. There could be a number of reasons for the remaining discrepancy between the experimental and theoretical ratios: (i) Our CHE and CHA results are not perfectly converged with respect to the angular averaging as can be seen in Fig.~\ref{fig:lebedev_convergence}, especially for the higher harmonic orders. (ii) The individual orbitals do not have the same symmetries as the full electron density, and we would therefore in principle need to include calculations for additional angles when we reconstruct the total dipole moment from the individual orbitals. (iii) Our calculations do not include macroscopic effects such as phase matching or focal volume averaging. Such calculations are not currently within computational reach if one wants to include the TDDFT-calculated individual-molecule response, but it is interesting to contemplate whether (for example) the different intensity dependence of the three yields could make a meaningful difference in the yield ratios. (iv) The use of the LDA-ADSIC exchange-correlation functional could possibly be limiting the accuracy of the calculations, although this is still an open question \cite{vanleeuwen1994, mack2013}. 
                               
Nevertheless, we are encouraged by the success of the simple ADK-based correction of the HHG yields as we have shown here, and further studies of how to most properly implement such a scaling could be fruitful. A simple scaling is especially appealing given that grid-based TDDFT calculations of HHG are relatively accessible computationally, and lend themselves well to systematic studies, despite their well-known tendency to lead to incorrect ionization potentials \cite{casida2000, lemierre2005, mack2013}.    


\section{Summary and Discussion}

We have presented a quantitative comparison of the orientation-averaged HHG yields driven by an MIR laser field in three different ring-shaped organic molecules, calculated using grid-based TDDFT with an LDA+ADSIC exchange-correlation functional. The orientation averaging was done using the Lebedev quadrature scheme, and we exploited the dynamical symmetries of the molecule in the linearly polarized driving field in order to sharply reduce, by 60-80\%, the number of computations necessary for the orientation-averaged spectrum. 

We compared the calculated yields to experimental yields as documented in \cite{alharbi2015}. We found that the raw yield ratios for BNZ versus CHE compared reasonably well to experimental results, with the BNZ yield moderately larger than the CHE yield. In contrast, we found that the raw yield for CHA was much too large, and that this could be attributed to contributions from lower-lying orbitals (HOMO-2 and below) whose ionization potentials were strongly underestimated in our DFT and  TDDFT calculation. We then implemented  a simple orbital-resolved scaling of the yields with a ADK-rate-based correction factor, which corresponds to adjusting the ionization step of the HHG process according to the ratio corresponding to the computed ADK rates of the experimental and theoretical ionization potentials. This resulted in substantially better agreement between the calculated and measured yield ratios, and in particular in a strong suppression of the CHA yield compared to the other two molecules. 

The correction factor implemented in these results is in the spirit of the three-step model and its QRS implementation, namely that the HHG yield can be separated into a contribution from ionization and a contribution from rescattering. In addition, we are implicitly assuming that the ionization rate out of a particular orbital itself has two independent contributions: a contribution from simple tunnel ionization that can be corrected, as well as a contribution that depends on the shape of the orbital. Our results showed an improved agreement between the relative yields between the three molecules studied here, and we are optimistic that further studies into a simple orbital-based correction factor would be fruitful, especially given how TDDFT-based computations are currently the only viable approach to calculating HHG spectra in large molecules. 

{\textbf{Data Availability}}: The data and Python scripts used to generate the figures in this paper are available upon reasonable request.

\section*{Acknowledgements}

This work was supported by the U.S.\ Department of Energy, Office of Science, Office of Basic Energy Sciences, under Award No.~DE-SC0012462.
Portions of this research were conducted with high performance computational resources provided by Louisiana State University (\url{http://www.hpc.lsu.edu}) and the Louisiana Optical Network Infrastructure (\url{http://www.loni.org}).

\appendix

\section*{Appendix A: Angular Averaging with Lebedev Quadrature}

\setlength{\tabcolsep}{4pt}
\renewcommand{\arraystretch}{1.25}
\begin{table}[t]
    \centering
    \begin{tabular}{?{0.5mm}C?{0.5mm}C|C|C|C?{0.5mm}}
        \cline{2-5}
        \multicolumn{1}{c?{0.5mm}}{} & \multicolumn{4}{c?{0.5mm}}{Number of Simulations}\\
        \hline
        $k$ & total & BNZ & CHE & CHA\\
        \hline \hline
        3 & 6 & 3 & 3 & 3\\
        \hline
        5 & 14 & 4 & 5 & 5\\
        \hline
        7 & 26 & 7 & 9 & 9\\
        \hline
        11 & 50 & 10 & 15 & 15\\
        \hline
    \end{tabular}
    \caption{Number of points in the Lebedev sampling of $\mathbb{S}^{2}$ of order $k$, and the number of distinct Octopus simulations required once we have taken advantage of the dynamical symmetries in BNZ, CHE, and CHA.}
    \label{tab:num_sims}
\end{table}

Table~\ref{tab:num_sims} shows the number of unique simulations required for calculating the orientation-averaged HHG spectrum for a Lebedev grid of order $k$ \cite{lebedev1975}, once accounting for the dynamical symmetries of each molecule. BNZ, which has the highest symmetry, goes from 50 total points in the Lebedev-11 grid to 10 unique simulations, a reduction by a factor of five. CHE and CHA have roughly the same degree of symmetry, and see an approximately three-fold reduction in the number of simulations needed -- though we note that the orientations used in each calculation are not necessarily the same for both molecules.

We now briefly describe how to exploit the dynamical symmetries \cite{ceccherini2001, bunker2006} in the molecule, when working on a Cartesian grid. Since the Lebedev quadrature has octahedral-rotation symmetry, there are many pairs of orientations within the Lebedev grid that are identical up to a $\pi$ rotation around one of the three Cartesian axis. Thus, we define $\pi$-rotation operators around the three Cartesian axes:
\begin{equation}
    \begin{cases}
        \hat{\Pi}_{x} \rho(\vec{r}, t) = \rho(x, -y, -z, t)\\
        \hat{\Pi}_{y} \rho(\vec{r}, t) = \rho(-x, y, -z, t + T_{L}/2)\\
        \hat{\Pi}_{z} \rho(\vec{r}, t) = \rho(-x, -y, z, t + T_{L}/2)
    \end{cases}
\end{equation}
where the laser is polarized along the $x$-axis, and the time translation used in the $\hat{\Pi}_{y}$ and $\hat{\Pi}_{z}$ operators is because after a $\pi$ rotation around the $y$ or $z$ axes, the electric field is reversed relative to the molecule, and $E(t) = -E(t + T_{L}/2)$ for a monochromatic electric field.

As an example, suppose that the ground-state densities corresponding to two different orientations of the molecule are identical after one has been rotated by $\pi$ radians around the $z$-axis; in mathematical terms, 
\begin{equation}
    \begin{aligned}
        \rho[\theta^{\prime}, \phi^{\prime}](\vec{r}, t) &= \hat{\Pi}_{z}\, \rho[\theta, \phi](\vec{r}, t)\\
        &= \rho[\theta, \phi](-x, -y, z, t + T_{L}/2).
    \end{aligned}
\end{equation}
We can then compute the dipole corresponding to $\rho[\theta^{\prime}, \phi^{\prime}](\vec{r}, t)$ in terms of $\rho[\theta, \phi](\vec{r}, t)$. In the $x$-direction,
\begin{align}
    \mu_{x}[\theta^{\prime}, \phi^{\prime}](t) &= \iiint_{\mathbb{R}^{3}} x\, \rho[\theta^{\prime}, \phi^{\prime}](x, y, z, t)\, \mathrm{d}^{3}\vec{r}\\
    &= \iiint_{\mathbb{R}^{3}} x\, \rho[\theta, \phi](-x, -y, z, t + T_{L}/2)\, \mathrm{d}^{3}\vec{r} \notag
\end{align}
The integral over $x$ gives a factor of -1 because of the multiplication with $x$. This is missing in the integral over $y$, which therefore gives a factor of +1. The integral over $z$ is unchanged. Finally, the time translation gives an extra factor of -1. Therefore, $\mu_{x}[\theta^{\prime}, \phi^{\prime}](t) = \mu_{x}[\theta, \phi](t)$ in this case. 

Repeating this process for the other two components of the dipole moment and the other two rotation operators, we find that 
\begin{equation}
    \begin{cases}
        \hat{1}: \vec{\mu}[\theta^{\prime}, \phi^{\prime}](t) = \vec{\mu}[\theta, \phi](t) \circ (+1, +1, +1)\\
        \hat{\Pi}_{x}: \vec{\mu}[\theta^{\prime}, \phi^{\prime}](t) = \vec{\mu}[\theta, \phi](t) \circ (+1, -1, -1)\\
        \hat{\Pi}_{y}: \vec{\mu}[\theta^{\prime}, \phi^{\prime}](t) = \vec{\mu}[\theta, \phi](t) \circ (+1, -1, +1)\\
        \hat{\Pi}_{z}: \vec{\mu}[\theta^{\prime}, \phi^{\prime}](t) = \vec{\mu}[\theta, \phi](t) \circ (+1, +1, -1)
    \end{cases}
\end{equation}
where $\circ$ denotes the Hadamard product (component-wise multiplication). Eq.~{9} has been checked against actual dipole accelerations computed in Octopus to ensure their validity.

\section*{Appendix B: Agreement between Dipole Signals}

\begin{figure}[!t]
    \centering
    \includegraphics[width=\linewidth]{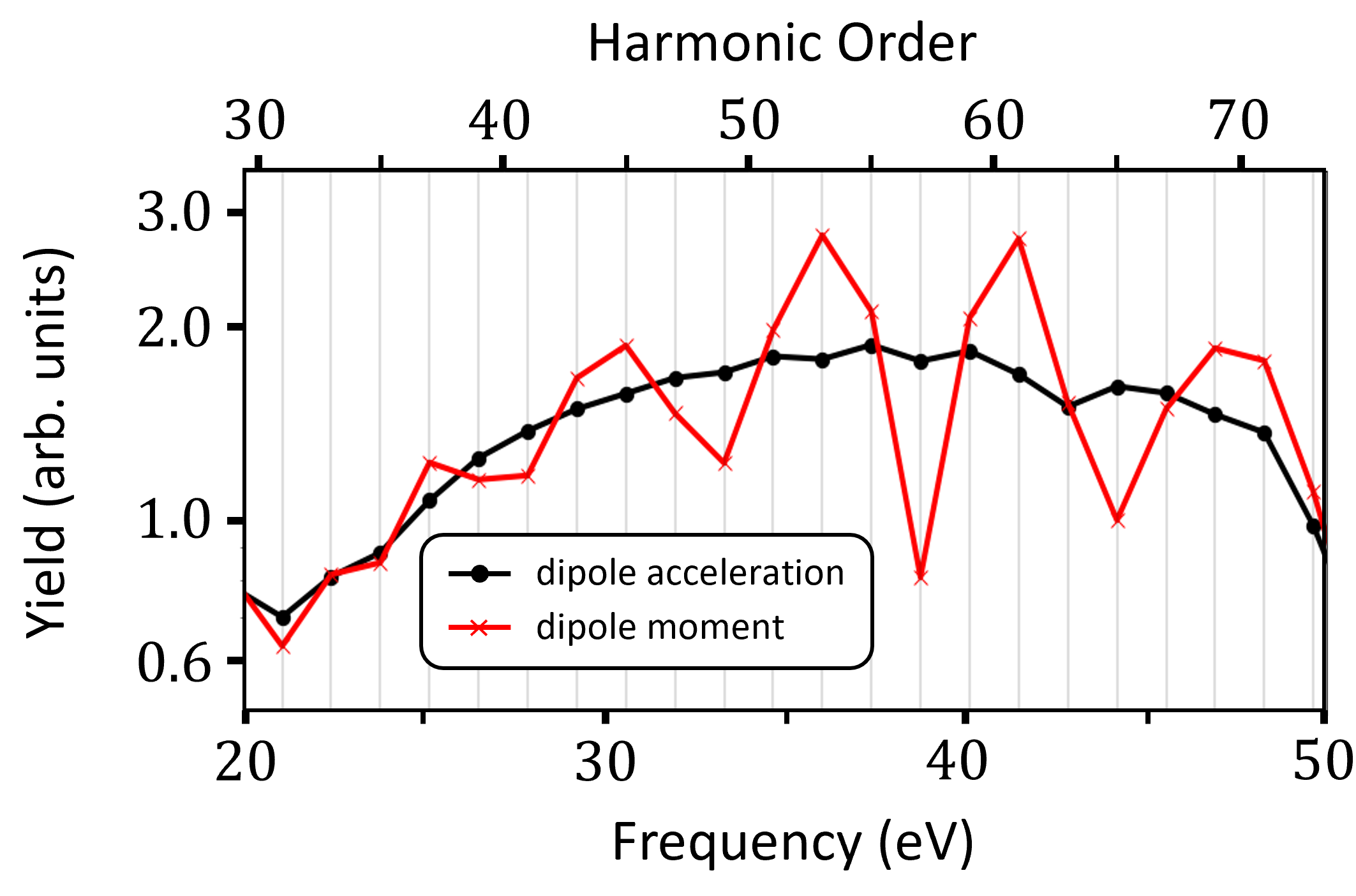}
    \caption{Comparison between different dipole signals used to compute the HHG yield in BNZ: the total dipole acceleration (black) and the total dipole moment (red).}
    \label{fig:dipole_comparison}
\end{figure}

Figure~\ref{fig:dipole_comparison} compares the orientation-averaged HHG spectra (using the Lebedev 11 grid) computed in two different ways, for a BNZ molecule irradiated by a 45~TW/cm$^{2}$ laser pulse. The spectrum computed from the windowed Fourier transform of the dipole acceleration $a(t) = \tfrac{\mathrm{d}^{2}}{\mathrm{d}t^{2}} \mu(t)$ (black) -- recall Eq.~{\ref{eq:spectrum}} -- is quite smooth over the entire harmonic plateau, while the spectrum computed from the windowed Fourier transform of the dipole moment $\mu(t)$ (red) is not. The oscillations in the red curve are caused by an interference between two or more dipole signals from different orientations, and reflects that the calculation of the dipole moment is in general less stable than that of the acceleration and leads to noisier spectra. Apart from the oscillations, the two spectra agree very well in their shape, and in the average yield. 

We note here that the spectrum that results from the sum of the five highest-lying orbitals (as is shown in Fig.~\ref{fig:corrected_spectra}, after correcting for the ADK rates) is smaller than the full spectrum from all the orbitals by about a factor of two. This is true for all three molecules, and therefore does not affect the ratios between the yields.  We interpret this as field-driven population transfer between orbitals during the HHG process, which results in a small amount of recombination to orbitals that were not initially tunnel ionized. 

\setlength{\tabcolsep}{4pt}
\renewcommand{\arraystretch}{1.25}
\begin{table}[t]
    \centering
    \begin{tabular}{?{0.5mm}c?{0.5mm}C|C|C?{0.5mm}}
        \cline{2-4}
        \multicolumn{1}{c?{0.5mm}}{} & \multicolumn{3}{c?{0.5mm}}{Vertical $I_{p}$ (TDDFT)}\\
        \hline
        Orbital & BNZ & CHE & CHA\\
        \hline \hline
        HOMO & \multirow{2}*{9.83} & 9.12 & 10.58\\
        \cline{1-1} \cline{3-4}
        HOMO-1 & & 10.87 & 10.50\\
        \hline
        HOMO-2 & \multirow{2}*{11.70} & 10.37 & 10.52\\
        \cline{1-1} \cline{3-4}
        HOMO-3 & & 10.77 & \multirow{2}*{11.50}\\
        \cline{1-3}
        HOMO-4 & 12.50 & 12.10 & \\
        \hline
    \end{tabular}
    \caption{Calculated vertical ionization potentials for the five highest-occupied Kohn-Sham orbitals in BNZ, CHE, and CHA.}
    \label{tab:vIPs}
\end{table}

\section*{Appendix C: Rescaling with the Vertical Ionization Potential}

\begin{figure}[!t]
    \centering
    \includegraphics[width=\linewidth]{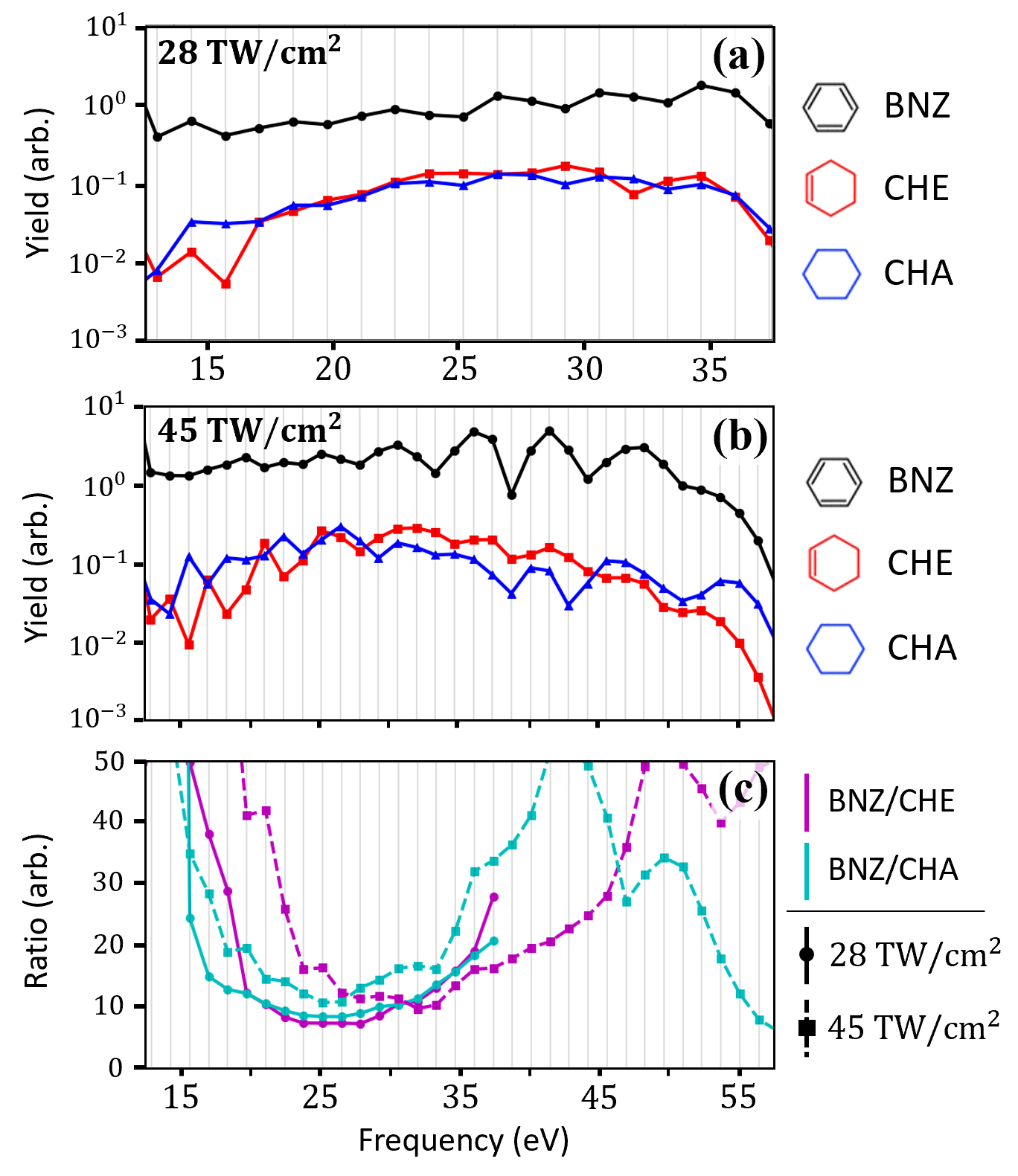}
    \caption{(a,b) Orientation-averaged HHG yields with the contributions from the highest lying orbitals weighted using the vertical ionization potentials computed in Octopus, for two different intensities. (c) ADK-corrected HHG yield ratios between BNZ and either CHE or CHA, for both intensities.}
    \label{fig:vertical_IP}
\end{figure}

The ADK-corrected results shown in Figs.~\ref{fig:corrected_spectra} and \ref{fig:ratios} were calculating using the orbital energies as $I_p^{\text{thy}}$. However, one might argue that a more direct comparison to the experimental $I_p$ if \cite{kimura1981} would be to calculate the vertical $I_p$ for each orbital. Fig.~\ref{fig:vertical_IP} shows the corrected spectra for the two intensities, and the resulting yield ratios, when using calculated vertical $I_p$'s in Eq.~\ref{eq:adk_ratio} instead.

We compute the vertical $I_p$ of a particular molecular orbital by removing one electron from that orbital (using the \texttt{Occupations} block in Octopus) and then performing a ground-state DFT calculation on the remaining cation. The vertical ionization potential is then the energy difference between the total DFT energy of the neutral vs. the cation ground state. Note that for cyclohexane, we needed to use an eigensolver (\texttt{Eigensolver = evolution}) that is different from the default eigensolver in Octopus (\texttt{Eigensolver = cg}) in order to get some cations to converge. The calculated vertical $I_{p}$ values for the same orbitals we showed in Table I are shown in Table III.

Similar to Fig.~{\ref{fig:corrected_spectra}}, Figures~\ref{fig:vertical_IP}~(a) and (b) show the HHG spectra in BNZ, CHE, and CHA, using the vertical ionization potentials tabulated in Table~{\ref{tab:ips}}. Again, the BNZ spectrum (black) has been enhanced by the ADK correction factor, relative to the other two spectra, similar to the experimental results \cite{alharbi2015}. However, the CHE and CHA spectra are on top of one another, suggesting that CHA was not suppressed enough using the vertical-$I_p$ re-weighting scheme. Also, in Fig.~{\ref{fig:vertical_IP}}(c), we plot the ratios of the HHG spectra, similar to Fig.~{\ref{fig:ratios}}, but without the experimental ratios. Again, we have smoothed the ratios using a moving average over five points in order to remove the oscillations in the dipole signal seen most prominently in Fig.~{\ref{fig:dipole_comparison}}. Within the harmonic plateau, both the BNZ:CHE and BNZ:CHA ratios are between 10 and 30.

\end{document}